\documentclass[12pt,preprint]{aastex}

\begin{document}

\def\br{{\mathbf r}}
\def\bs{{\mathbf s}}
\def\bgrad{{\mathbf\nabla}}
\def\mmm{\left<\mu_m\right>}
\def\mmd{\left<\mu_2\right>}
\def\mmq{\left<\mu_4\right>}
\def\g{\gamma}
\def\esf{poo}

\title{Analytic expressions for mean magnification by a quadrupole
gravitational lens}

\author{
Tehani K.\ Finch\altaffilmark{1,2},
Lisa P.\ Carlivati\altaffilmark{3},
Joshua N.\ Winn\altaffilmark{4},
Paul L.\ Schechter\altaffilmark{1}
}

\altaffiltext{1}{Massachusetts Institute of Technology,
77 Mass.\ Ave., Cambridge, MA 02139; {\tt schech@achernar.mit.edu}}

\altaffiltext{2}{Howard University, Thirkield Hall, 2355 6th St.\ NW,
Washington, DC 20059; {\tt tfinch@howard.edu}}

\altaffiltext{3}{Harvard College, University Hall, Cambridge, MA
02138}

\altaffiltext{4}{Harvard-Smithsonian Center for Astrophysics, 60
Garden St., Cambridge, MA 02138; {\tt jwinn@cfa.harvard.edu}}

\begin{abstract}

We derive an analytic expression for the mean magnification due to
strong gravitational lensing, using a simple lens model, a singular
isothermal sphere embedded in an external shear field.  We compute
separate expressions for 2-image and 4-image lensing.  For 4-image
lensing, the mean magnification takes a particularly simple form:
$\mmq=\frac{3.6}{\g\left(1-\g^2\right)}$, where $\g$ is the external
shear.  We compare our analytic results to a numerical evaluation of
the full magnification distribution.  The results can be used to
understand the magnification bias that favors the discovery of
four-image systems over two-image systems in flux-limited lens
surveys.

\end{abstract}

\section{Introduction}

There have been several systematic surveys for examples of
multiple-image gravitational lensing of extragalactic sources, both at
optical wavelengths \cite[e.g.,][]{maoz93,surdej93,gregg01} and radio
wavelengths \citep{burke93,king99,winn00,browne02}.  The
interpretation of the number of lensed sources that are detected in a
flux-limited survey depends critically on the factors by which the
sources are magnified.  Lensed objects are drawn from a parent
population with smaller fluxes than their unlensed counterparts.  One
therefore samples a fainter portion of the source luminosity function
and a more distant range of source redshifts.  This well-known
``magnification bias'' also affects the relative frequencies of
different lens morphologies.  Four-image lenses are more highly
magnified than two-image lenses, and are subject to a stronger
magnification bias.

Early studies of lens statistics used lens models with circular
symmetry, such as singular isothermal spheres, for which the
magnification bias is easy to calculate \cite[e.g.,][]{turner84}.
However, symmetric models only produce 2-image lenses and cannot be
used to interpret the statistics of each morphology separately.  More
recently, numerical studies have been performed with more realistic
elliptical lens models
\cite[e.g.,][]{king96,wallington93,keeton97,rusin01}.  While numerical
calculations are certainly valuable, they tend to obscure the roles of
competing contributions to what is being calculated.  Analytic forms
for such things as luminosity functions and magnifications may be no
more accurate, but they produce results that are more readily
understood.

With this motivation we examined a number of simple elliptical lens
models and found one for which the mean magnification can be computed
analytically, for both 2-image and 4-image lensing.  The model is a
singular isothermal sphere in an external shear field (SIS+XS), which
is physically motivated and widely used.

In the next section we briefly review the mathematics and the
qualitative features of lensing by non-circular potentials.  In
\S~\ref{sec:calculation}, we specialize to the SIS+XS and derive the
analytic expressions for all of the relevant quantities, including the
mean magnification.  Finally, in \S~\ref{sec:numerical} we compare our
result to the full numerical evaluation of the magnification
distribution, and remark on applications to the interpretation of
lensing statistics.

\section{Review of lensing by elliptical potentials}
\label{sec:theory}

Lensing theory is well developed and has been expounded in detail by
\citet{schneider92}, among others.  The particular problem of lensing
by elongated potentials has been lucidly described by
\citet{blandford86}.  Here we briefly review only those concepts that
are required to follow the calculations presented in
\S~\ref{sec:calculation}.

We use the conventional definition of the lens potential,
\begin{equation}
\psi(r,\theta) \equiv \frac{2}{c^2} \frac{D_{ls}}{D_l D_s}
\int_0^{\infty} \Phi(D_{l}\br,z) dz,
\label{eq:lens-potential} 
\end{equation}
in which $\Phi$ is the Newtonian gravitational potential of the
galaxy; $\br$ is the sky-plane angular displacement; $z$ is the
line-of-sight coordinate; and $D_l$, $D_s$, and $D_{ls}$ are the
angular-diameter distances to the lens, to the source, and between
lens and source, respectively.  With this definition, the
correspondence between source position $\bs$ and image position $\br$
is
\begin{equation}
\bs = \br - \bgrad_{\br}\psi(\br).
\end{equation}
\label{eq:lens-equation}
The inverse magnification of a particular image $i$ is given by
\begin{equation}
\mu_i^{-1} = \left(1-\frac{\partial^2\psi}{\partial x^2} \right)
             \left(1-\frac{\partial^2\psi}{\partial y^2} \right) -
	            \left(\frac{\partial^2\psi}{\partial x \partial y}\right)^2,
\label{eq:magnification}
\end{equation}
which is negative when the image and source have opposite parity.  For
a survey in which sources are selected by total flux, the relevant
quantity is the total magnification $\mu_t(\bs)=\sum |\mu_i(\bs)|$,
the sum of the unsigned magnifications of all images of a source
located at $\bs$.

\begin{figure}
\figurenum{1}
\plotone{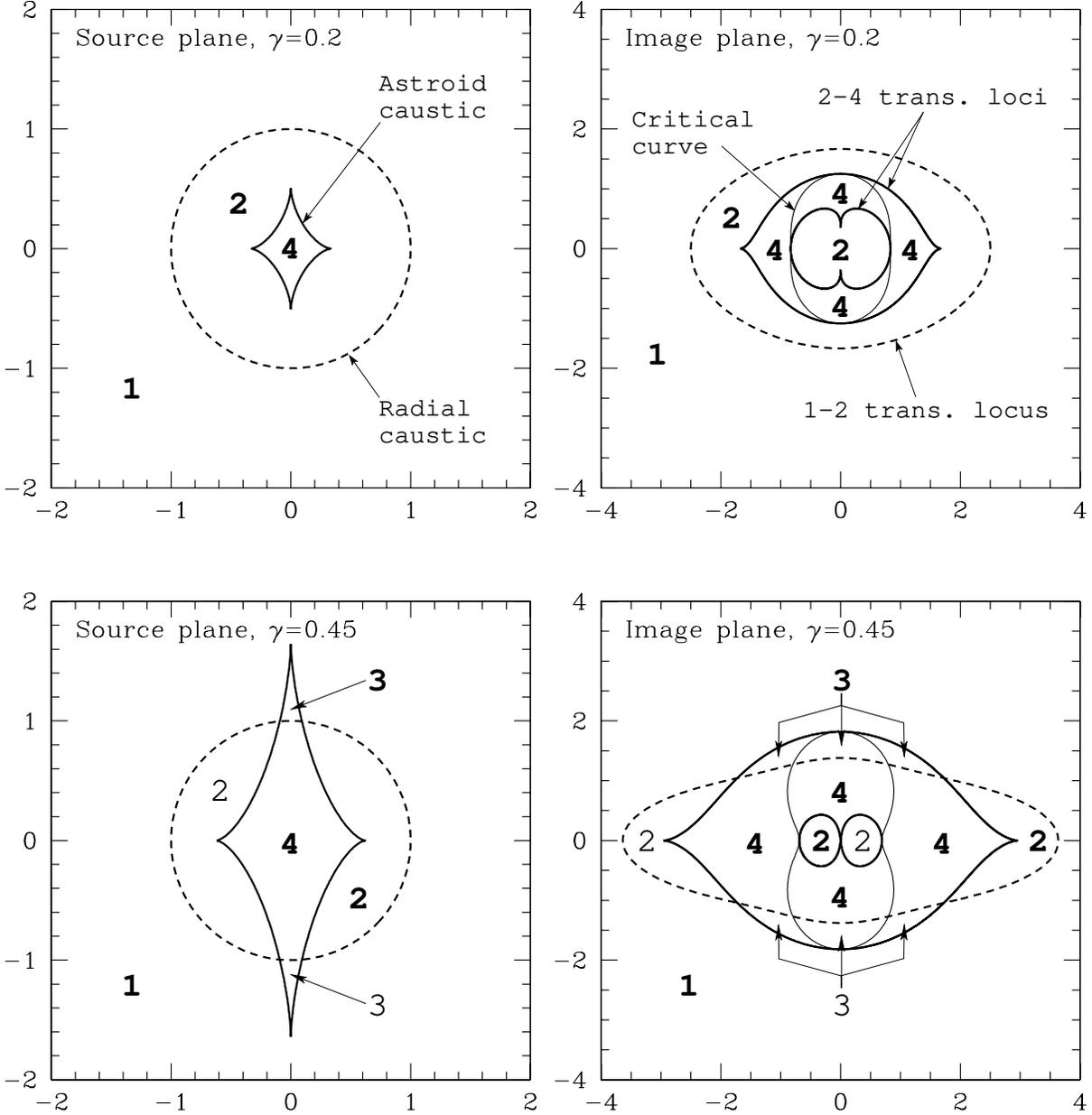}
\caption{ {\bf Top}.  Source plane (left) and image plane (right) for
the SIS+XS with $b=1$ and $\g=0.2$.  The curves discussed in
\S~\ref{sec:theory} are labeled.  The regions are labeled by image
multiplicity, e.g., a point source in a region of the source plane
marked ``4'' corresponds to four points in each of the four regions of
the image plane marked ``4.''
{\bf Bottom}. Same, but for $\g=0.45$.
}
\label{fig:sisxs}
\end{figure}

Most of the observed galaxy lenses can be described (at least
qualitatively) using centrally concentrated, singular potentials.  If
the isopotential contours are only moderately elliptical, such
potentials produce either 1, 2, or 4 images of a background source, in
locations that can be understood with reference to the curves plotted
in the top two panels of Figure~\ref{fig:sisxs}.  The curves are
plotted for the SIS+XS, but the following discussion applies to many
singular elliptical potentials.\footnote{Non-singular lens potentials
produce an odd number of images.  In such cases, the image plane has
an additional critical curve and transition locus surrounding the
origin.}

For a source outside the {\em radial caustic}\footnote{Strictly
speaking, this curve is not a caustic, because the potential is
singular.  For this reason, the curve has also been designated the
radial {\em pseudo-caustic} \citep{evans98} or {\em cut}
\citep{kormann94}.}, one image is produced outside the 1--2 {\em
transition locus}.  A source on the radial caustic corresponds to one
image on the 1--2 transition locus and another image at the origin,
with zero flux.  If the source moves from the radial caustic to the
{\em astroid caustic}, the outer image moves to the outer 2--4
transition locus, while the inner image acquires a nonzero flux and
moves to the inner 2--4 transition locus.  On the astroid caustic,
three images are produced: one on each of the 2--4 transition loci;
and one on the {\em critical curve}, with infinite flux.  Inside the
astroid caustic, the latter image splits into two images with finite
flux, and the result is one image in each of the regions marked ``4''
in Figure~\ref{fig:sisxs}.  The caustics and critical curve are well
known to those who study gravitational lensing, but the transition
loci receive less attention than they deserve---in particular, they
are crucial for our calculation of mean magnification.

For more elongated potentials, two cusps of the astroid caustic pierce
the radial caustic and are often referred to as ``naked cusps.''  The
relevant curves in the source plane and the image plane are shown in
the bottom panels of Figure~\ref{fig:sisxs}.  A source inside a naked
cusp corresponds to three images located on the same side of the
origin as the source---one in each of the three regions marked ``3''
in Figure~\ref{fig:sisxs}.

If the source position is random, the cross section for producing 4
images, $\sigma_4$, is the total area within the 4-image region of the
source plane (marked ``4'' in Figure~\ref{fig:sisxs}).  The 2-image
cross-section is the area of the 2-image region, and the 3-image
cross-section is the area of the 3-image region.  The cross-section
for multiple imaging is $\sigma_m = \sigma_2 + \sigma_3 + \sigma_4$.

The mean magnification for multiple-image lensing is the average value
of the total magnification for all source positions that result in
more than one image:
\begin{equation}
\mmm = \frac{\int \mu_t(\bs) d\bs}{\int d\bs} =
       \frac{\int_{\hbox{\footnotesize mult}} d\br}{\int d\bs},
\label{eq:mean-magnification}
\end{equation}
where the integral of $d\br$ extends over the region in the image
plane in which multiple images are found.
Eq.~\ref{eq:mean-magnification} holds because the Jacobian for the
$\bs\rightarrow\br$ transformation is equal to the magnification.
Alternatively, one can visualize the calculation as follows.  Because
lensing conserves surface brightness, a source consisting of uniform
brightness in the region formed by the union of the caustics will
result in an image of uniform brightness within the union of the
transition loci.  The mean magnification for multiple-image lensing,
which we denote $\mmm$, is the ratio of the image area to the source
area. The mean magnification of 4-image lenses, denoted $\mmq$, is the
ratio of 4-image region of image space and the 4-image region of
source space.  The mean magnification of 2-image and 3-image lenses
can be visualized in a similar manner.

Of the approximately 100 observed cases of multiple-image
gravitational lensing by galaxies, almost none require naked cusps for
explanation.  The one possible exception is APM~08279+5255
\citep{lewis02}.  Because naked-cusp systems are apparently rare in
nature, and because the calculations described below cannot be done
analytically for naked-cusp systems, in this paper most of our
attention will be on systems in which the astroid caustic is interior
to the radial caustic.

In that case, to find the cross-sections and mean magnifications for
2-image and 4-image lensing, it is sufficient to calculate the areas
enclosed by the radial caustic ($A_r$), astroid caustic ($A_a$), the
1--2 transition locus ($A_t$), and the inner and outer 2--4 transition
loci ($A_i$ and $A_o$). As long as the astroid caustic is contained
entirely within the radial caustic, we have:
\begin{eqnarray}
\sigma_4 & = & A_d \nonumber\\
\sigma_2 & = & A_r-A_d \nonumber\\
\mmq     & = & \frac{A_o-A_i}{A_d}\nonumber \\
\mmd     & = & \frac{A_t-A_o+A_i}{A_r-A_d}
\label{eq:xsec-and-mag}
\end{eqnarray}

\section{Analytic expressions for the SIS+XS}
\label{sec:calculation}

We concentrate on the case of a singular isothermal sphere in an
external shear field (SIS+XS), for which the potential is
\begin{equation}
\psi(r,\theta) = br + \frac{\g r^2}{2} \cos 2\theta.
\label{eq:sisxs}
\end{equation}

Here $(r,\theta)$ are polar coordinates in the image plane, $b$ is the
Einstein ring radius (which sets the mass scale of the lens), and $\g$
is the magnitude of the external shear.  The SIS is an idealization of
a spherical dark-matter halo with a flat rotation curve, and the
external shear is a non-circular perturbation.  This potential can be
thought of as a truncated multipole expansion of the lens potential
\cite[see, e.g.,][]{kovner87,schneider91,trotter00}.  The SIS+XS has
been used extensively in the literature to model observed lenses,
often as a starting point for more complicated models \cite[see,
e.g.,][]{kochanek91,keeton97,lehar97,schechter97}.

By computing the second derivatives of this potential
and using Eq.~\ref{eq:magnification}, we obtain the magnification
\begin{equation}
\mu^{-1} = 1 - \g^2 - \frac{b}{r}
\left(1 - \g\cos 2\theta \right).
\label{eq:magnification-sisxs}
\end{equation}

We now derive expressions for all the relevant curves and the areas
enclosed by them.  We identify the radial caustic $(x_r, y_r)$ by
requiring that it map to the origin under Eq.~\ref{eq:lens-equation},
obtaining
\begin{eqnarray}
x_r = -b\cos\theta \nonumber \\
y_r = -b\sin\theta,
\label{eq:radial caustic}
\end{eqnarray}
which is a circle of radius $b$, enclosing an area $A_r = \pi b^2$.

Next, we identify the 1--2 transition locus by reverse-mapping the
radial caustic back into the image plane, and finding the non-zero
solution $r_t(\theta_t)$:
\begin{eqnarray}
x_r = -b\cos\theta =
\left(1-\g\right)r_t\cos\theta_t - b\cos\theta_t \nonumber \\
y_r = -b\sin\theta =
\left(1+\g\right)r_t\sin\theta_t - b\sin\theta_t.
\label{eq:sdtc-1}
\end{eqnarray}
We eliminate the parameter $\theta$ by squaring and adding the
equations, leading to the solution
\begin{equation}
r_t(\theta_t) = 2b \left(
\frac{1 - \g\cos 2\theta_t}
{1 - 2\g\cos 2 \theta_t + \g^2} \right).
\label{eq:sdtc-2}
\end{equation}
The area enclosed by the 1--2 transition locus can be expressed in
terms of definite integrals that are analytically tractable:
\begin{eqnarray}
A_t & = & \frac{1}{2} \int_0^{2\pi} r_t^2 d\theta_t \nonumber \\
    & = & 8b^2 \int_0^{2\pi}
           \left[ \frac{1}{1 - 2\g\cos 2\theta + \g^2} -
                  \frac{\g^2 (1 - \cos 4\theta)}{(1 - 2\g\cos 2\theta + \g^2)^2} \right] d\theta \\
    & = & 4\pi b^2 \left( \frac{1 - \frac{\g^2}{2}}{1 - \g^2} \right). \nonumber
\label{eq:stdc-area}
\end{eqnarray}

To identify the critical curve $r_c(\theta_c)$, we require the inverse
magnification $\mu^{-1}$ given by Eq.~\ref{eq:magnification-sisxs} to
be zero, obtaining
\begin{equation}
r_c(\theta_c) = \frac{ b(1 - \g\cos 2\theta_c) }{1 - \g^2}.
\label{eq:critical-curve}
\end{equation}
We obtain parametric equations for the astroid caustic $(x_a,y_a)$ by
requiring the astroid caustic to map to the critical curve under the
lens mapping, Eq.~\ref{eq:lens-equation}:
\begin{eqnarray}
x_a & = & -\frac{2b\g}{1 + \g} \cos^3 \theta_c \nonumber \\
y_a & = &  \frac{2b\g}{1 - \g} \sin^3 \theta_c.
\label{eq:astroid-caustic}
\end{eqnarray}
The maximum value of $x_a$ is $x_* = \frac{2b\g}{1 + \g}$, and the
area within the astroid caustic is
\begin{equation}
A_a = 4 \left| \int_0^{x_*} y_a dx_a \right| =
      \frac{16b^2\g^2}{1-\g^2}\int_0^{\frac{\pi}{2}} 3\sin^4\theta \cos^2\theta d\theta =
      \frac{3\pi b^2}{2} \frac{\g^2}{1 - \g^2}.
\label{eq:astroid-caustic-area}
\end{equation}
We note that the astroid caustic is completely interior to the radial
caustic for $\g<\frac{1}{3}$.

Under the lens mapping, the astroid caustic has 2 images apart from
the critical curve: the inner 2--4 transition locus, $r_i(\theta_i)$,
and the outer 2--4 transition locus, $r_o(\theta_o)$.  In order to
solve for these transition loci, which we refer to generically as
$r_4(\theta_4)$, we insert the parametric
equations~\ref{eq:astroid-caustic} into Eq.~\ref{eq:lens-equation}:
\begin{eqnarray}
-\frac{2b\g}{1 + \g} \cos^3 \theta_c & = &
\left(1-\g\right)r_4\cos\theta_4 - b\cos\theta_4 \nonumber \\
 \frac{2b\g}{1 - \g} \sin^3 \theta_c & = &
\left(1+\g\right)r_4\sin\theta_4 - b\sin\theta_4.
\label{eq:transition-loci-1}
\end{eqnarray}
After eliminating $r_4$ from this pair of equations, and defining
$t\equiv\cos\theta_c$, we obtain
\begin{equation}
(\cos\theta_4 - t)^2
\left[ \cos^2\theta_4 + 2t(1-t^2)\cos\theta_4 - t^4 \right] = 0,
\label{eq:transition-loci-2}
\end{equation}
which has three solutions: the double root
$\cos\theta_4=t$ corresponding to the critical curve, and
the 2--4 transition loci
\begin{equation}
\cos\theta_{i,o} =
t \left( t^2 - 1 \pm \sqrt{t^4 - t^2 + 1} \right).
\label{eq:transition-loci-3}
\end{equation}
We insert these solutions for $\cos\theta_4$ into the first expression
of Eqs.~\ref{eq:transition-loci-1} to obtain parametric equations for
the 2--4 transition loci:
\begin{equation}
r_4(t) = \frac{b}{1-\g}
\left[
1 - \frac{2\g t^3}{(1+\g)\cos\theta_4(t)}
\right].
\label{eq:transition-loci-4}
\end{equation}
It is possible to find the area within each locus separately---but for
the calculation of mean magnification, the key quantity is the area
between the 2--4 transition loci, which is (after lengthy algebra)
\begin{equation}
A_o-A_i
    = 4\times \frac{1}{2}\int_0^{\frac{\pi}{2}} (r_o^2 - r_i^2)d\theta
    = \frac{4\pi b^2\g}{(1-\g^2)^2}
      (I_1 + I_2\g),
\label{eq:transition-loci-area-1}
\end{equation}
where $I_1$ and $I_2$ are defined as
\begin{eqnarray}
I_1 & = & \frac{2}{\pi}\int_0^{\frac{\pi}{2}}
    \frac{t_i^3(\theta)-t_o^3(\theta)}{\cos\theta}d\theta \nonumber \\
I_2 & = & \frac{2}{\pi}\int_0^{\frac{\pi}{2}}
    \left[ \frac{t_i^3(\theta)-t_o^3(\theta)}{\cos\theta}
         + \frac{t_i^6(\theta)-t_o^6(\theta)}{\cos^2\theta} \right] d\theta,
\label{eq:integrals}
\end{eqnarray}
and $t_{i,o}(\theta)$ are the inverse relations of
Eq.~\ref{eq:transition-loci-3}.  It is straightforward to prove
$I_2=0$ by demonstrating that the integrand is antisymmetric about
$\theta=\frac{\pi}{4}$ \cite[see][]{finch01}, and $I_1$ is a
dimensionless number independent of $b$ and $\g$.  Numerically
we find $I_1=1.35111$, leading to the result
\begin{equation}
A_o - A_i = \frac{4\pi b^2 I_1\g}{(1-\g^2)^2} \approx
\frac{17b^2\g}{(1-\g^2)^2}.
\label{eq:transition-loci-area-2}
\end{equation}

Now that all of the relevant curves and areas have been calculated, we
compute cross sections and mean magnifications using
Eqs.~\ref{eq:xsec-and-mag}.  We emphasize that these results are only
valid for $\g<\frac{1}{3}$, when the astroid caustic is completely
interior to the radial caustic.  To find mean magnifications for
$\g>\frac{1}{3}$, one must determine the intersection points of the
various curves, for which we do not have analytic solutions.  In
\S~\ref{sec:numerical} we present numerical results for this case.

The SIS+XS is the only example we have found that has all the
qualitative features desired of a model for an elliptical potential
and for which the areas bounded by the transition loci are all
analytic.  An example of another commonly-used potential for which we
do not have analytic formulas for all the areas is
\begin{equation}
\psi(r,\theta) = br + \g br\cos 2\theta,
\label{eq:self-similar}
\end{equation}
which can be thought of as a lowest-order multipole expansion of the
singular isothermal ellipsoid (SIE) density distribution \cite[see,
e.g.,][]{kormann94}.  The astroid caustic is contained completely
within the radial caustic for $\g < \frac{1}{5}$.  For this potential
we found analytic expressions for all the curves except the 1--2
transition locus, using a procedure similar to the one described
above.  We were therefore unable to compute $\mmd$, but all the other
quantities are listed in Table~\ref{tbl:results}.

\begin{deluxetable}{cccccc}
\tablecaption{Cross sections and mean magnifications\label{tbl:results}}
\tablewidth{0pt}

\tablehead{
\colhead{$\psi(r,\theta)$} &
\colhead{$\g_{\hbox{max}}$} &
\colhead{$\sigma_2$} &
\colhead{$\sigma_4$} &
\colhead{$\mmd$} &
\colhead{$\mmq$} 
}

\startdata

$br + \frac{\g r^2}{2} \cos 2\theta$ &
$\frac{1}{3}$ &
$\pi b^2\left( \frac{1-\frac{5}{2}\g^2}{1-\g^2} \right) $ &
$\frac{3\pi b^2}{2} \frac{\g^2}{1-\g^2} $ &
$4\times \frac{ \left(1-\frac{\g^2}{2}\right)\left(1-\g^2\right) - I_1\g }
              { \left(1-\frac{5\g^2}{2}\right)\left(1-\g^2\right) } $ &
$\frac{8I_1}{3\g(1-\g^2)}$ \\

$br + \g br\cos 2\theta$ &
$\frac{1}{5}$ &
$\pi b^2 \left( 1 - \frac{15}{2}\g^2 \right) $ &
$6\pi b^2 \g^2$ &
\nodata &
$\frac{4I_1}{3\g}$ \\ 

\enddata

\tablenotetext{*}{The number $I_1$ is defined by
Eq.~\ref{eq:integrals}, and has the approximate value 1.35111.}

\end{deluxetable}

\section{Discussion}
\label{sec:numerical}

\begin{figure}
\figurenum{2}
\plotone{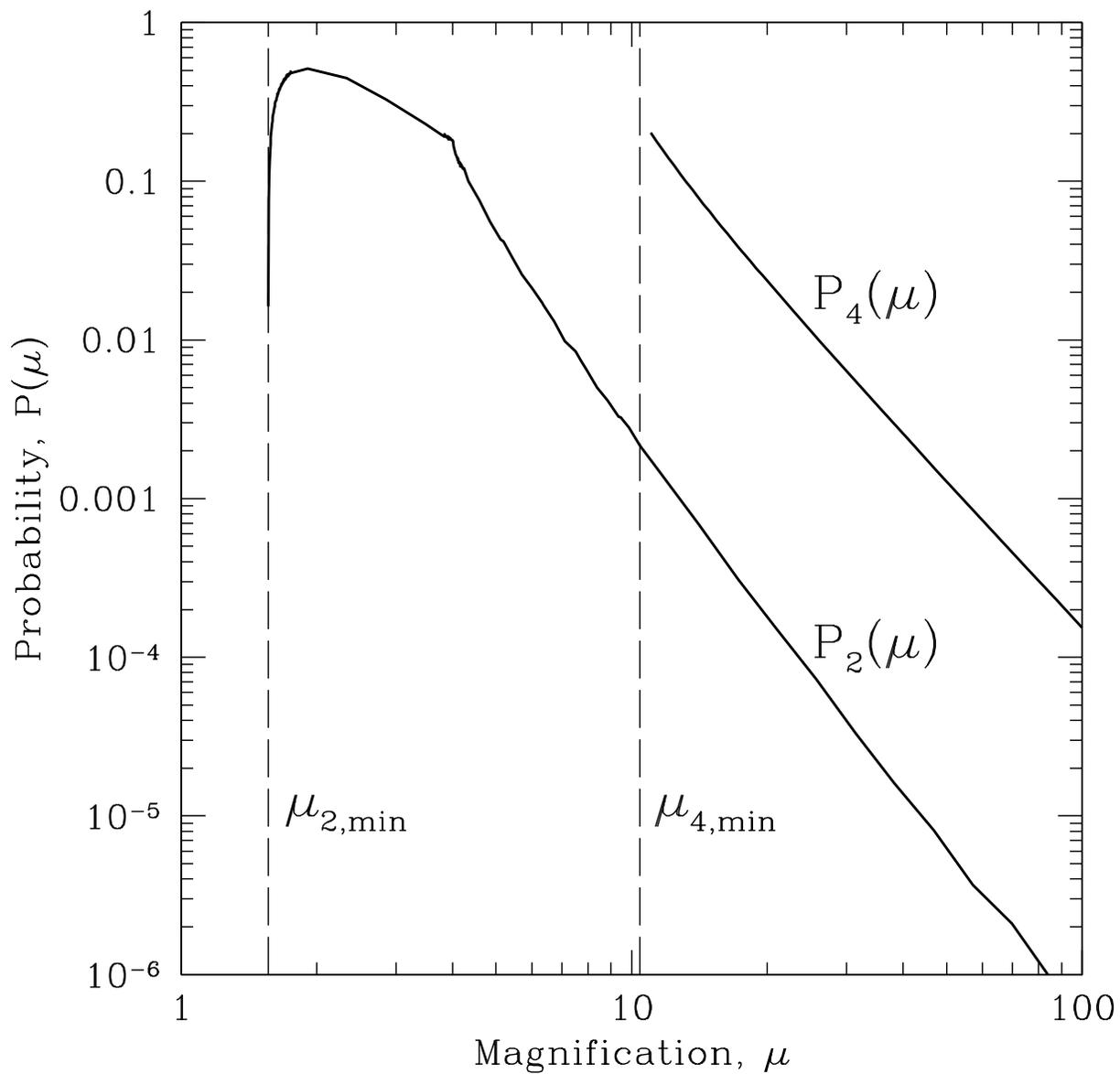}
\caption{ Probability distribution of magnification by the SIS+XS
model, shown here for the case $\g=0.2$.  The 2-image distribution
($P_2$) and the 4-image distribution ($P_4$) are plotted
separately. The minimum magnifications, given by
Eqs.~\ref{eq:minimum-magnification-2}
and~\ref{eq:minimum-magnification-4}, are marked.  }
\label{fig:probs}
\end{figure}

To put the analytic results in context, we numerically computed the
full magnification distribution for multiple imaging by the SIS+XS,
denoted $P_m(\mu$).  We parameterized the points within the radial
caustic and computed the total magnification of each point, by
inverting Eq.~\ref{eq:lens-equation} to find the images and
Eq.~\ref{eq:magnification-sisxs} to derive the magnifications.  We
performed this procedure separately for the 2-image and 4-image
regions of the source plane, so that the magnification distribution
for each morphology ($P_2$ and $P_4$) could be computed separately.
The results are plotted in Fig.~\ref{fig:probs} for the case $\g=0.2$.

The general features of the distribution can be readily understood.
The minimum magnification occurs when the source is on the $x$-axis of
the radial caustic, giving
\begin{equation}
\mu_{2,\hbox{\footnotesize min}} = \frac{2}{\left(1+3\g\right) \left(1-\g\right)}.
\label{eq:minimum-magnification-2}
\end{equation}
There is a discontinuity at the minimum magnification for 4-image
systems, which occurs for a source at the origin:
\begin{equation}
\mu_{4,\hbox{\footnotesize min}} = \frac{2}{\g \left(1-\g^2\right)}.
\label{eq:minimum-magnification-4}
\end{equation}
The high magnification tail is produced by sources in the vicinity of
the astroid caustic.  It can be shown generically that the asymptotic
probability for caustic-dominated events varies as $\mu^{-3}$ (see,
e.g., Blandford \& Narayan 1987; or Schneider, Ehlers, \& Falco 1992,
p.\ 319), and indeed the logarithmic slope of $P_4$ as computed
numerically is $-3.1$.  The downward kink between
$\mu=\mu_{2,\hbox{\footnotesize min}}$ and $\mu_{4,\hbox{\footnotesize
min}}$ occurs at the minimum magnification for a source just outside
the astroid caustic.

The number of multiple-image lenses in a flux-limited sample can be
determined from the magnification distribution.  As described in
detail by \citet{turner84}, in a survey of objects with intrinsic
luminosity function $\phi(L,z)$ and a flux limit corresponding to an
intrinsic luminosity $L_{\hbox{\footnotesize lim}}$, the bias factor
$B$ by which multiple-image lenses at a given redshift $z$ are
over-represented is
\begin{equation}
B = \frac{
\int_{L_{\hbox{\tiny min}}}^{\infty} dL \int_{\mu_{\hbox{\tiny min}}}^{\infty} \mu^{-1}P_m(\mu) \phi(\frac{L}{\mu},z) d\mu
}
{
\int_{L_{\hbox{\tiny min}}}^{\infty} \phi(L,z) dL 
}.
\label{eq:magnification-bias}
\end{equation}
The bias factors for 2-image and 4-image lensing can also be computed
separately, by replacing $P_m$ with $P_2$ or $P_4$.  For the common
case that the luminosity function is approximated by a power law,
\begin{equation}
\phi(L,z) = \frac{\phi_\star(z)}{L_\star(z)} \times \left( \frac{L}{L_{\star}(z)} \right)^{-\beta},
\end{equation}
the bias factor reduces to $B=\left<\mu^{\beta-1}\right>$, which is
simply the mean magnification for $\beta=2$.  As pointed out by
\citet{rusin01}, the value $\beta=2$ happens to be a good
approximation for the radio source populations that have been searched
in recent lens surveys (JVAS, King et al.\ 1999; CLASS, Browne et al.\
2002; Winn et al.\ 2000).  Our analytic results are therefore directly
relevant to this case, in which previous calculations were performed
by numerical integration and Monte Carlo simulation.

For example, the JVAS and CLASS surveys have found roughly equal
numbers of 2-image and 4-image lenses, a result that various authors
have struggled to interpret because it seems to require a
magnification bias for 4-image lensing that is larger than expected
\citep{king96,keeton97,rusin01}.  A resolution considered by these
authors is that the matter distribution in the lens galaxies is
systematically more elongated than the observed light distribution of
elliptical galaxies.

We can estimate of the required value of $\g$ from our calculations of
mean magnification.  The ratio of 4-image lenses to 2-image lenses,
including the effect of magnification bias, is
\begin{equation}
\frac{N_4}{N_2} = \frac{\sigma_4}{\sigma_2} \times
\frac{B_4}{B_2}.
\end{equation}
For the SIS+XS model and a $\beta=2$ luminosity function, the result
is
\begin{equation}
\frac{N_4}{N_2} = \frac{\sigma_4\mmq}{\sigma_2\mmd} =
\frac{I_1\g}{ \left(1-\frac{\g^2}{2}\right) \left(1-\g^2\right) - I_1\g }.
\label{eq:quads-to-doubles}
\end{equation}
This result is rather specialized, depending as it does on the
assumption of a particular lens model and source luminosity function,
but it has the virtue of being analytic.  To obtain equal numbers of
4-image and 2-image systems we require $\g=0.32$, for which
$\sigma_2/\sigma_4$ and $B_4/B_2$ are both approximately 5.

\begin{figure}
\figurenum{3}
\plotone{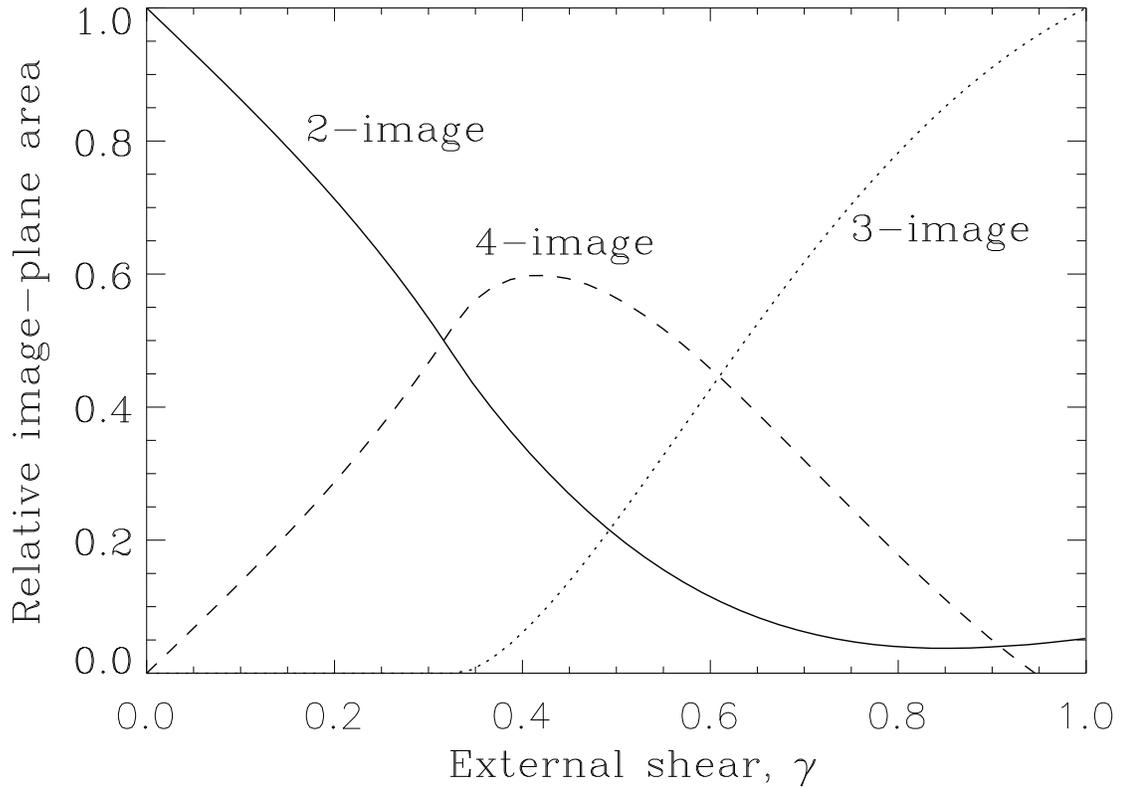}
\caption{ Numerical evaluation of the relative areas of the 2-image,
3-image, and 4-image regions of the image plane, for the SIS+XS model.
Under the assumptions described in \S~\ref{sec:numerical} these curves
also give the ratios of 2-image, 3-image, and 4-image systems expected
in a flux-limited survey. }
\label{fig:areas}
\end{figure}

For $\g>\frac{1}{3}$ one must also consider the possibility of 3-image
systems produced by naked cusps.  In this case the computation of
areas in the image plane must be done numerically.
Figure~\ref{fig:areas} shows the results of a numerical calculation of
the relative areas of the 2-image, 3-image, and 4-image regions of the
image plane, for $0<\g<1$.  As before, if $\beta=2$ then this plot can
also be interpreted as the expected fractions of 2-image, 3-image, and
4-image lenses discovered in a flux-limited survey.  For
$\g<\frac{1}{3}$ the results are identical to our analytic formula.
For $\g=\frac{1}{3}$, naked-cusp images are possible but occur with
vanishing probability; they constitute 10\% of the lenses at
$\g\approx 0.45$ and they dominate the sample for $\g>0.61$.

In more generic settings, the full magnification distribution is
required.  The 4-image magnification distribution, at least, can be
approximated by $P_4(\mu) \propto \mu^{-3}$ for $\mu >
\mu_{4,\hbox{\footnotesize min}}$.  The mean of this approximate
distribution is
\begin{equation}
\mmq \approx \frac{
\int_{\mu_{4,\hbox{\footnotesize min}}}^{\infty} \mu^{-2} d\mu
}{
\int_{\mu_{4,\hbox{\footnotesize min}}}^{\infty} \mu^{-3} d\mu
}
= \frac{4}{\g(1-\g^2)},
\label{eq:approximate-mean-magnification-4}
\end{equation}
which has the same dependence on $\g$ as the exact expression (see
Table~\ref{tbl:results}) and differs only by a constant fraction of
$4/3.6$ or 11\%.

\section{Conclusion}
\label{sec:conclusion}

We have described a geometrical method for computing the mean
magnification for multiple imaging by non-circular lens potentials.
For the special case of a singular isothermal sphere embedded in an
external shear field, we have derived analytic expressions for the
mean magnification for 2-image lensing and 4-image lensing.  The
results can be used, in certain circumstances, to compute the
magnification bias for 2-image and 4-image lensing.

\acknowledgements We thank Nick Morgan, Jackie Hewitt, and David Rusin
for helpful discussions.  We gratefully acknowledge the support of the
Graduate Student Office at MIT (T.K.F.), the Research Science
Institute at MIT (L.C.), US NSF grant AST-9616866 (P.L.S.) and an NSF
Astronomy \& Astrophysics Postdoctoral Fellowship (J.N.W.) under grant
no.\ 0104347.

\end{document}